# The Elusive High-T$_c$ Superinductor

Yogesh Kumar Srivastava[1,2], Manoj Gupta[1,2], Manukumara Manjappa[1,2], Piyush Agarwal[1,2], Jérôme Lesueur[3] and Ranjan Singh[1,2,*]

[1]*Division of Physics and Applied Physics, School of Physical and Mathematical Sciences, Nanyang Technological University, Singapore 637371, Singapore*

[2]*Centre for Disruptive Photonic Technologies, The Photonics Institute, Nanyang Technological University, Singapore 637371, Singapore*

[3]*Laboratoire de Physique et d'Etude des Matériaux, CNRS, ESPCI Paris, PSL Research University, UPMC, 75005, Paris, France*

[*]Email: ranjans@ntu.edu.sg

**Abstract:**

Ginzburg-Landau (*GL*) parameters formed the basis for Abrikosov's discovery of the quantum vortex of a supercurrent in type-II superconductor with a normal core of size ξ, the superconductor coherence length and circulating supercurrent induced magnetic field diverging as log(*1/r*) from the core with a decay length of the London penetration depth, $\lambda_L$. In 1964, J. Pearl predicted the slowly decaying (1/$r^2$) field around a vortex spreading out to Pearl length, $P_L = \frac{2\lambda_L^2}{t}$, in a superconductor film of thickness $t < \lambda_L$. However, his quintessential theory failed to predict the existence of giant kinetic inductance (GKI) that arises from the enlarged screening currents of the vortex. Here, we discover giant kinetic inductance in a high-T$_c$ metasurface due to the 1400% expansion of the vortex screening supercurrent from $\lambda_L$ to $14\lambda_L$ in ultrathin film meta-atom of $\lambda_L$/7 thickness, which leads to the emergence of terahertz superinductance possessing quantum impedance exceeding the resistance quantum limit of $R_Q = \frac{h}{(2e)^2} = 6.47\ k\Omega$ by 33%. Our discovery presents a new class of high-T$_c$ superconductor electronic, photonic, and quantum devices enabled through metasurface designed at the Pearl length scales, providing novel applications in quantum circuitry, metrology, and single photon kinetic inductance detectors.

**Introduction:**

The superinductors are the quantum circuit elements with zero DC resistance, low internal losses, and characteristic impedance exceeding the quantum resistance $R_Q = \frac{h}{(2e)^2} = 6.47\ k\Omega$.[1,2] The superinductors are also immune to charge offset fluctuations and hence vital for enabling compact electro-optical components for applications in quantum technologies and micro/nano-circuits.[3,4] However, implementation of superinductance is forbidden with pure geometric inductors and capacitors as the maximum impedance is limited to vacuum impedance. The superinductors could be realized using superconductors (SC), as they satisfy three key attributes, zero DC resistance, low losses, and large impedance within the desired frequency regime.[1] Most superinductors realized to date are based on either Josephson junctions or disordered materials to enhance kinetic inductance, complicating the device geometry and limiting their applications only to small footprints, extremely low temperatures and low operating frequencies.[5]

One possible way to realize superinductors is through engineering the Cooper-pair charge transport in ultrathin SCs. The charge transport by Cooper pairs in a SC is mainly described by the London equations. However, the current flow in SC strip of thickness $t$ lower than London penetration depth ($\lambda_L$) and width $w$ larger than $\lambda_L$, is non-trivial and could not be expressed using simple analytical solution of London equations.[6] In such strips, the current density is relatively uniform throughout the thickness of the film, while it decays across the width on moving away from the edges. W. A. Brower's first described this behavior using an approximate solution of the London equations assuming current density at a distance '$x$' from the center given by $J_x = J(0)e^{-[\frac{t(\frac{w}{2}-x)}{a\lambda_L^2}]}$, where '$a$' is the constant.[6] Further, experimental studies by Broom and Rhoderick also suggested elongated magnetic field distribution and current screening up to a length much larger than $\lambda_L$.[7] Later, independent studies were performed by J. Pearl in 1964 to characterize the decay profile of the magnetic field of a vortex in a Type-II SC with thickness lower than $\lambda_L$, which differs from the decay profile of magnetic field around an 'Abrikosov vortex'.[8-10] The magnetic field around the Abrikosov vortex falls off rapidly with distance, $r$, from the center of the core as log ($1/r$), whereas it decays much slower as $1/r^2$ around the Pearl vortex.[10] The spread of magnetic field around the core of the Pearl vortex in a thin SC is inversely proportional to the thickness, $t$, of the SC film and is quantified by a length scale called Pearl length, given as $P_L = \frac{2\lambda_L^2}{t}$.[10,11] This effect of Pearl vortex is analogous to the screening current in a thin SC strip, which could be quantified by length scale $P_L$. Therefore, $P_L$ of an ultrathin SC governs the Cooper pair transport that introduces loss due to the formation of large screening region within the superconductor and accounts for further enhancement in overall inductance of the material, that is given by the summation of *geometric inductance '$L_G$'* and *kinetic inductance '$L_K$'*. The geometric inductance of the system depends on the geometry of the current flow. In contrast, the kinetic energy of charge carriers that participate in the flow of current determines the $L_K$.[12-14] Therefore, at terahertz frequencies, where small size of the devices limit $L_G$, realizing giant enhancement of $L_K$ provides an attractive solution to realize compact superinductors.

Intrinsically $L_K$ depends on the scattering time ($\tau$), and DC conductivity ($\sigma_0$) of the material and is expressed as, $L_K = \frac{\tau}{\sigma_o} = \frac{m^*}{ne^2}$, where $m^*$, $n$ and $e$ are the effective mass, number density, and charge of the current-carrying carriers, respectively.[15] The material having heavier charge carriers with longer scattering time provides a sizeable $L_K$.[13,14,16] In metallic systems, $L_K$ is negligible at terahertz frequencies ($\omega \sim 10^{12}$ rad s$^{-1}$) due to the short scattering time of charge carriers ($\tau < 10^{-14}$ s), leading to dominant resistive component of the effective impedance overtaking the contribution from $L_K$. However, $L_K$ becomes significant for metallic systems at higher frequencies ($\omega \sim 10^{15}$ rad s$^{-1}$).[16] At terahertz and lower frequencies, SCs provide an excellent material platform for observing $L_K$ due to the relatively high $\tau$ and $m^*$ of the Cooper pairs.[12,13] In the presence of an alternating field, the heavier mass of the cooper pairs in SC results in the enormous kinetic energy of the carriers, that is invested in equivalent inductive energy stored in the magnetic field, thereby generating $L_K$. SC wires with the advantage of tuning the geometrical parameters provide a suitable platform for tailoring $L_K$. The $L_K$ of a SC strip of length $l$, width $w$, and thickness $t$ is given by $L_K = \frac{\mu_o \lambda_L^2 l}{wt} = \frac{m^*}{2n_s e^2}\left(\frac{l}{wt}\right)$, where $\mu_o$, $n_s$, $m^*$ and $2e$ are, respectively, the permeability of free space, number density, effective mass, and effective charge of the Cooper pairs.[17] Therefore, $L_K$ of the SC wires could be enhanced either by reducing thickness and width or by manipulating carrier density actively by introducing external perturbations. Most interestingly, formation of elongated screening current in the ultrathin SC forms non-superconducting regions at the edges of the strip and reduces the effective superconducting cross-section area ($w*t$) of the strip, momentarily enhancing the $L_K$.[10] The mathematical expression of $L_K$ due to current screening could be expressed in terms of *Pearl length* as $L_K = \mu_0 P_L(\frac{l}{2w})$. Therefore, controlling $P_L$ opens a favorable pathway for realizing superinductance either by varying geometrical parameters of the SC structure or by actively manipulating $n_s$ and $\lambda_L$.

Here, we report the first observation of superinductance in sub-London penetration depth thick high-$T_c$ Yttrium Barium Copper Oxide (YBCO) superconductor Pearl metadevice at terahertz frequencies powered by the giant kinetic inductance (GKI). We observe a record temperature-dependent resonance frequency shift of 290 GHz (44%) in 25 nm thin Pearl metadevice. We also developed a transmission line circuit model for the microscopic understanding of high-frequency physics, impedance characterization, and experimental observation of the onset of superinductance caused by the Pearl length in the ultrathin SCs.[11,18]. The large impedance of ~ 8.6 kΩ, (greater than the quantum resistance $R_Q$) in 25 nm thick Pearl metadevice culminates the realization of superinductance without the need for Josephson junction design. Further, upon illuminating the Pearl metadevice with an optical pulse, we observe ultrasensitive dynamic optical control of the Pearl length, resulting in a controllable $L_K$ and a record time-dependent frequency agility of 76 GHz at an ultrafast timescale of 7 ps. Our work provides the novel pathway to design meta-atoms at the superconducting Pearl length scale to give birth to high-$T_c$ superinductance at terahertz frequencies, boosting the prospects of enabling high-frequency micro/nano circuits for applications in quantum electronics, communication devices, kinetic inductance based single photon detectors and radiation sensors.[19]

**Result and discussion:**

The schematic of thin current-carrying SC wires of thickness smaller than $\lambda_L$ and width comparable to or larger than the Pearl length is depicted in Fig. 1(a). The current in both the wires is uniform in thickness while decaying across the width of the wires, as shown by the color map from pink (max current) to blue (zero current) region. The current screens a larger cross-section of rectangular wire with narrow width $w'$ compared to the wire of width $w''$. The screening currents penetrate up to a length scale $P_L$ in both wires, as shown by black double-headed dotted arrows. A magnetic flux region is formed due to the screening currents and hinders the transport of Cooper pairs that critically govern $L_K$ in the sub-$\lambda_L$ thickness of SC. To validate the sub-$\lambda_L$ nature of SC strips, we first characterize ultrathin films of high-$T_c$ YBCO with thickness 25 nm and 50 nm deposited on a *r*-cut sapphire substrate of thickness 500 μm. Even though it is critical to use extremely thin YBCO to achieve large $P_L$ and $L_K$, fabricating YBCO films of thickness lower than 25 nm drastically diminishes the superconductor transition temperature ($T_c$) from ~85 K to a lower value.[20] Therefore, we restrict our investigation to the minimum 25 nm thickness of YBCO to investigate the role of Pearl length in enabling superinductance.

Fig. 1(b) shows the measured temperature-dependent complex conductivity of 25 nm thick YBCO film at 0.6 THz, which is the resonance frequency (shown in Fig. 2(a)) of the corresponding Pearl metadevice at 7 K. The real part of the complex conductivity shows a slow enhancement as the temperature reduces from 300 K to 70 K and reduces gradually on further decreasing the temperature to 7 K. While the imaginary part of the conductivity shows a steep increase by several orders of magnitude as the sample temperature reaches below the transition temperature $T_c = 81$ K, showing the onset of superconductivity in the YBCO film. Complete details of the terahertz transmission spectra and frequency dependent complex conductivity are given in section S3 of the supplementary information (SI). The frequency dependent complex conductivity is fitted using two-fluid model of superconductivity to obtain normal carrier density ($n_n$), and Cooper pair density ($n_s$) of 25 nm and 50 nm thick YBCO at different temperatures, as shown in Fig. S3 (d) and (e).[21,22] Using calculated $n_s$ values, $\lambda_L$ for 25 nm and 50 nm YBCO films are obtained using equation $\lambda_L = \sqrt{\frac{m}{2\mu_o n_s e^2}}$, where *m* and *e* are the mass and charge of the electron. We further analyze temperature-dependent $\lambda_L$ by fitting it with the following parametric equation,

$$\lambda_L(T) = \frac{\lambda_0}{\sqrt{1-\left(\frac{T}{T_c}\right)^\alpha}} \qquad (1)$$

where $\lambda_0$ is the penetration depth at T=0K, and α is the exponent determining temperature dependence.[23,24] The scattered and dashed lines in Fig. 1(c) depict the measured values of $\lambda_L$ and the fitting curves, respectively, while the extracted fitting parameters are given in Table S1 of the SI. The parametric fitting reveals $\lambda_0$ values of 172.7 nm and 143 nm for 25 nm and 50 nm YBCO films, respectively, which agrees with the reported values in the literature.[23] The $T_c$ of 25 nm and 50 nm YBCO films were measured to be 81 K and 84 K, respectively. Evidently, 25 nm and 50 nm YBCO films possess sub-$\lambda_L$ thicknesses and show uniform current distribution in the corresponding SC strip. Further, to realize the effect of current flow from such thin strips of SC, we designed and fabricated typical inductive-capacitive (LC) resonant metamaterial structures over a large area of 1 cm × 1 cm, using photolithography and wet chemical etching methods. The detailed fabrication process is given in Methods section. Fig. 1(d) depicts the optical image of fabricated metadevice and the inset shows geometrical dimensions of its unit cell.

Experimentally measured temperature-dependent transmission spectra of 25 nm YBCO Pearl metadevice are shown in Fig. 2(a) and they depict the remarkable redshift of 262 GHz as the sample temperature increased from 7 K to 80 K, much higher compared to the previously reported terahertz metadevices.[25,26] To further justify the record frequency-shift obtained in the 25 nm (thinnest) Pearl metadevice, we compare the temperature-dependent frequency-shift of 25 nm thin sample with 50 nm YBCO metadevice with identical geometrical parameters resulting in a 165 GHz frequency shift, as shown in Fig. 2(b). The observed frequency shift of 262 GHz and 165 GHz in 25 nm and 50 nm Pearl metadevices, respectively, correspond to the 44% and 19% frequency agility, with respect to the central LC resonance frequency of 0.62 THz and 0.85 THz measured at lowest temperature of 7 K. The record redshift in the resonance frequency for the 25 nm-Pearl metadevice is attributed to the onset of GKI due to the enlarged screening currents in the ultrathin SC-strips.

To understand the origin of this enhanced temperature-dependent frequency agility and high-frequency behavior of Pearl metadevice, we developed an equivalent transmission line circuit model for the designed structures consisting of total inductance ($L_T = L_G + L_K$), capacitance ($C$), and resistive terms ($R_a$, $R_L$, and $R_r$). The input impedance ($Z_{in}$) of the structure in terms of circuit elements is given by $Z_{in} = \frac{1}{\left(\frac{1}{R_r}\right) + \left(\frac{1}{R_L + j\omega L}\right) + (j\omega C)}$. The circuit's schematic is revealed in the inset of Fig. 2(d), while the detailed circuit and fitting method is provided in Section 4 of SI. The measured transmission spectra of 25 nm and 50 nm YBCO Pearl metadevice at each temperature are fitted using the circuit model to estimate the *total inductance* ($L_T$). We further analyze the calculated temperature-dependent $L_T$ by fitting an analytical equation of inductance for a thin rectangular SC wire, where $L_T$ as a function of temperature $T$ could be expressed as

$$L_T = L_G + \left(\frac{\mu_0 l}{tw}\right)\left(\frac{{\lambda_0}^2}{1-\left(\frac{T}{T_c}\right)^\alpha}\right) \qquad (2)$$

where $L_G$ represents the geometric inductance. The detailed fitting parameters are revealed in Table S2 of SI. The $L_K$ of 25 nm and 50 nm Pearl metadevice has been obtained by subtracting $L_G$ of the corresponding meta-structure, as shown in Fig. 2(c). It is noted that $L_K$ of 25 nm and 50 nm Pearl metadevice increases with rising temperature and is primarily responsible for large temperature-dependent frequency tuning. However, contrary to the theoretical calculations, $L_K$ does not diverge but shows saturation just below the $T_c$. This peculiar effect arises at higher temperatures because the resistive component of complex impedance dominates over the inductive component and leads to saturation in $L_K$.[25] Moreover, while comparing the $L_K$ observed by Pearl metadevices, we note a surprising enhancement of $L_K$ only in the 25 nm YBCO sample, while 50 nm YBCO device does not reflect this phenomenon. Further, Fig. 2(d) reveals that the $Z_{in}$ of 50 nm metadevice, is 5.6 kΩ ($<R_Q$), whereas, for 25 nm YBCO Pearl-metadevice the $Z_{in}$ ~8.6 kΩ ($>R_Q$) at 7 K, showing compelling evidence of the first terahertz superinductor.

To better understand the origin of the onset of GKI in ultrathin superconductors, we calculate the temperature-dependent Pearl length of identical unstructured YBCO films using the measured $\lambda_L$ (as shown in Fig. 1(c)). The $P_L$ of 25 nm and 50 nm thick YBCO film is found to be 2.26 μm and 0.97 μm at 6 K, which is about 14$\lambda_L$ and 6$\lambda_L$, respectively, and increases monotonically with increasing temperature, as shown in Fig. 2(c). The temperature at which $P_L$ exceeds $w$ that leads to a huge enhancement in $L_K$ is called the *onset temperature* ($T_K$) of GKI. Notably, in the case of 25 nm YBCO, $P_L$ exceeds the resonator arm width $w$ (4 μm) at $T_K$ =55 K (shown by yellow shaded region in Fig. 2c), triggering the onset of the GKI. However, in the case of 50 nm YBCO, $P_L$ grows equal to or larger than 4 μm at 76 K, which is close to $T_c$ of the film. The overlapping of $T_K$ with the $T_c$ of YBCO superconducting phase transition limits the observation of GKI for the 50 nm thick Pearl-metadevice. Therefore, by investigating Pearl metadevices of two different thicknesses, we experimentally validate our hypothesis and achieve Pearl length controlled GKI and superinductance.

To further strengthen the observation of Pearl length enabled superinductor, we probe the dependence of GKI onset temperature ($T_K$) on the SC strip-width ($w$) in Pearl metadevice of thickness 25 nm. For this study, we carefully chose three Pearl metadevices with strip-widths $w$ of 2.8 μm, 4 μm, and 7 μm. First, we performed simulations using a 3D MLSI (Multilayer Superconducting Integrated Circuits) simulation tool based on the numerical solutions involving

London and Maxwell equations and monitor the expansion of the surface currents in the metadevice unit cell at 7 K.[27] We observed uniform current distribution in the vertical arms of all the structures, as shown in Fig. 3(a) ( i, ii and iii). However, in the case of $w$ = 2.8 µm and 4 µm, the counter circulating loop currents (top and bottom loop of the resonator strip) along the horizontal strips cancel with each other within the Pearl length as seen through the dark shaded regions. In the case of $w$ = 7 µm, the width is large enough to accommodate the circulating currents excited in the common strip for two adjoining loops, and therefore, they show weak influence of Pearl length at low temperatures. We further validate our results by experimentally capturing LC resonance frequency shift with respect to the resonance frequency at the lowest temperature of 7 K for fabricated Pearl metadevice with $w$ = 2.8, 4, and 7 µm, as shown in Fig. 3(b). The Pearl metadevice with the largest $w$ = 7 µm shows a record resonance frequency tuning of 290 GHz, followed by 262 GHz and 168 GHz for $w$ = 4 and 2.8 µm, respectively. Further, we used transmission line circuit model to fit the measured transmission spectra of all the samples at each temperature to estimate $L_T$ and $Z_{in}$ observed in the Pearl metadevices. Calculated $L_T$ and $Z_{in}$ of Pearl metadevices with $w$ = 2.8, 4, and 7 µm, are shown by black, red, and blue scatter plots, respectively, in Fig. 3(d) and Fig. S6 of SI, respectively. It is observed that the $L_T$ of the Pearl metadevice decreases with increasing $w$ due to the larger cross-sectional area. The $L_T$ of all three samples increases with temperature but shows different enhancement trends due to the varying Pearl length enabled screening current in the strips of the metadevices. On subtracting $L_G$ from $L_T$, we obtained GKI values of, $L_K$ = 564 pH, 478 pH and 80 pH close to T$_c$ for $w$ = 2.8, 4, and 7 µm, respectively. The $Z_{in}$ for the sample with $w$ = 2.8 and 4 µm is found to be 8.4 kΩ (at 7 K) and 8.6 kΩ (at 7 K), respectively, which are larger than theoretical quantum impedance ($R_Q$ = 6.47 kΩ), showing the successful realization of superinductance. However, for the sample with $w$ = 7 µm, the $Z_{in}$ of 2.4 kΩ (< $R_Q$) was observed. We further analyze the measured temperature-dependent inductance of each sample through analytical fitting with Equation 2. On careful analysis, we observed good fits of temperature-dependent $L_T$ for temperatures below 34 K, 55 K, and 68 K for $w$ = 2.8, 4, and 7 µm, respectively, as shown by the green curves in Fig. 3(d). The extracted fitting parameters have been given in Table S2 of SI. The fitting result for each case turns out to be $\lambda_0$ = 180 nm and $\alpha$ = 2, respectively, which are close to the measured values of $\lambda_0$ = 172.7 nm and $\alpha$ = 2.5 for 25 nm YBCO film. The deviation in $\lambda_0$ and $\alpha$ could arise due to the lithography process-induced changes in ultrathin YBCO. The theoretical effective length $l$ obtained from the fitting for all three devices was larger than the geometrical dimensions of the single resonator loop due to the complex geometrical design and current flow in metadevice. The fitted curve deviates from the measured values when fitting the data in the extended temperature range above 34 K, 55 K, and 68 K for $w$ = 2.8, 4, and 7 μm, respectively. We extrapolated the fitted curves (black dashed curves) up to a larger temperature range using identical fitting parameters in Fig. 3 (d) to highlight this deviation from measured values. Interestingly, these temperatures where we observe the deviation from the measured values match the onset temperature ($T_K$) of GKI for each of the metadevices, as shown by vertical dashed lines in Fig. 3(c). Further, we plot the mean-error value between the theoretical values obtained from the fitted curve and experimentally measured $L_T$ values in Fig. 3(e). The existence of non-zero difference values above $T_K$ = 34 K, 54 K, and 64 K clearly verifies Pearl length transpired onset of GKI in ultrathin SC Pearl metadevices.

By using the optical-pump terahertz probe spectroscopy (OPTP), we further demonstrate active and ultrafast control of the Pearl length ($P_L$) and GKI ($L_K$) in the 25nm thick Pearl-metadevice.[28] Here, we irradiated 25 nm thick YBCO Pearl metadevice and YBCO thin-film with an optical pulse of wavelength 800 nm and measured the large dynamic change in the conductivity in the form of differential terahertz transmission by varying the time delay ($\tau_p$) between optical pulse and terahertz probe. The Cooper pair dynamics captured for pump fluence of 52 μJ/cm$^2$ and at temperature of 7 K show two distinct peaks in the dynamics, as shown by the blue dashed curve in Fig. 4(b). The dual peak behavior arises from Cooper pair dissociation due to incident optical beam and etalon from the back surface of the substrate.[28] Fig. 4(a) depicts the transmission spectra of 25 nm thick Pearl metadevice measured at various $\tau_p$. Surprisingly, upon optical irradiation, the Pearl metadevice depicts large amplitude modulation and significant resonance frequency agility of 76 GHz within a time interval of 7.2 ps, shown by the shift between the black (0 ps) and red curves (7.2 ps) in Fig. 4(a). The blue curve in Fig. 4(a) depicts that the metadevice resonance tends to reappear on further increasing time delay due to the restoration of superconductivity. We further highlight that the dynamic resonance frequency of 25 nm Pearl metadevice closely follows the Cooper dynamics (see Fig. 4 (b)), which signifies the active-control of both cooper-pair transport and the superinductance properties in the Pearl-metadevice. The red dots in Fig 4(b) represents the LC resonance frequency shift of 25 nm YBCO metadevice measured at time delay $\tau_p$ corresponding to 0 ps time delay. To obtain more insights of the processes leading to this large transient frequency tuning, we calculated the Pearl length of the 25 nm YBCO film upon identical optical irradiation (fluence = 52 µJ/cm$^2$, temp. = 7 K) at different $\tau_p$. From Fig. 4(c), it is evident that the measured $P_L$ exhibits large dynamic change from its equilibrium value of 2.38 µm to 3.26 µm at the maximum dissociation point upon irradiation with an optical pump pulse of relatively low fluence within 7.2 ps. More exciting consequence of this dynamic change in $P_L$ has been shown in Fig. 4(d), which shows the corresponding time-dependent variation in the $L_K$ of 25 nm Pearl metadevice measured under identical conditions. The $L_K$ here depicts about 54% transient modulation from its equilibrium value of 79 pH to 122 pH (7 K) at the peak of the dynamics compared to the 37% transient change in Pearl length that arises from the large surface currents in resonant Pearl metadevice resulting in melting charge orders.[29] Therefore, our results introduce a novel method to dynamically tune the kinetic inductance and superinductance in the ultrathin Pearl metadevice by light induced optical control of the Pearl length in the YBCO metadevice.

To conclude, we have experimentally demonstrated the first terahertz high-T$_c$ superinductor using a 25 nm thin YBCO metasurface demonstrating a record frequency-shift of 298 GHz, giant kinetic inductance of 564 pH, and characteristic impedance of 8.6 kΩ (>$R_Q$). We established that several exciting effects transpire from the Pearl length, which creates an extended screening area in the structured superconductor metasurfaces. Designing metadevices with geometrical dimensions comparable to Pearl length results in the fundamental deviation from the theory leading to the realization of giant kinetic inductance and superinductance. Furthermore, we demonstrate the active light induced switching of the high-T$_c$ superinductance at ultrafast timescales of 7 ps using extremely low fluence of optical excitation. These terahertz superinductors would find direct applications in the development of superconductor quantum circuits, information processing, ultra-compact inductors, metrology devices, and kinetic inductance-based broadband radiation detectors including terahertz single photon detection.

**Methods**

**a) Device Fabrication:** The superconductor metamaterial devices were fabricated using conventional UV illumination-based photolithography technique on commercially available YBCO films of size 1 cm x 1 cm and thickness 25 nm and 50 nm deposited on *r*-cut sapphire substrates. These 25 nm and 50 nm YBCO films have superconductor phase transition temperatures (T$_c$) of 81.6 K and 85.1 K and critical current density (J$_c$) of 0.7 and 2.3 MA/cm$^2$, respectively. The geometrical dimensions of the fabricated sample are shown in the inset of Fig.

1(d) of the main manuscript. First, a positive photoresist of thickness 1.5 µm was spin-coated on top of cleaned YBCO films and prebaked at 105 °C for 1 minute. These films were illuminated with UV light after mask alignment. The UV-exposed films were developed in a developer solution to get the desired pattern. The developed films were post-baked at a temperature of 110 °C for 2 minutes to harden the photoresist mask. The remaining photoresist acts as an etching protective layer. The undesired YBCO was wet etched using a 0.04% nitric acid solution followed by rinsing into the DI water. Finally, the desired samples were obtained by removing the photoresist using acetone. An optical image of the fabricated samples has been shown in Fig. 1(d).

**b) Terahertz Transmission and Optical Pump Terahertz Probe Measurements:** The terahertz transmission measurements of the YBCO metamaterial devices and thin films were carried out using a ZnTe based optical pump terahertz probe (OPTP) spectroscopy setup. The system was incorporated with a continuous flow liquid helium cryostat to facilitate measurements at cryogenic temperatures (up to 5 K). A pulsed laser beam of wavelength 800 nm (repetition rate 1 kHz, pulse width 35 fs) from a Ti: Sapphire amplifier laser was split into three parts. The first part of the beam has been used for the generation of terahertz pulse, the second for terahertz detection, and the third as an external pump beam for the photoexcitation of the superconducting samples. The optical pulse used for photoexcitation has a photon energy of 1.55 eV, which is much larger than the binding energy of the Cooper pairs (20-30 meV) present in the YBCO superconductor. The uniform excitation of the sample was ensured by illuminating an optical beam of diameter of 7 mm on the sample, which is much larger than the terahertz probe beam of diameter 4 mm. For the dynamic terahertz measurements, the delay between optical pump and terahertz probe beam was controlled by a translational delay stage. The dynamic change in the conductivity resulting due to the dissociation and formation of cooper pairs was measured in the form of differential terahertz transmission ($\frac{\Delta T}{T}$) by varying the time delay ($\tau_p$) between optical pulse and terahertz probe.

The transmission amplitude was estimated as the ratio of the transmitted terahertz electric field through the metamaterial device/thin film and reference (bare substrate) and is given by $\tilde{T}(\omega) = \frac{\tilde{E}_S(\omega)}{\tilde{E}_R(\omega)}$, where $\tilde{E}_S(\omega)$ and $\tilde{E}_R(\omega)$ are the Fourier transform of the transmitted electric field through the sample and reference, respectively.

**Figure 1:**

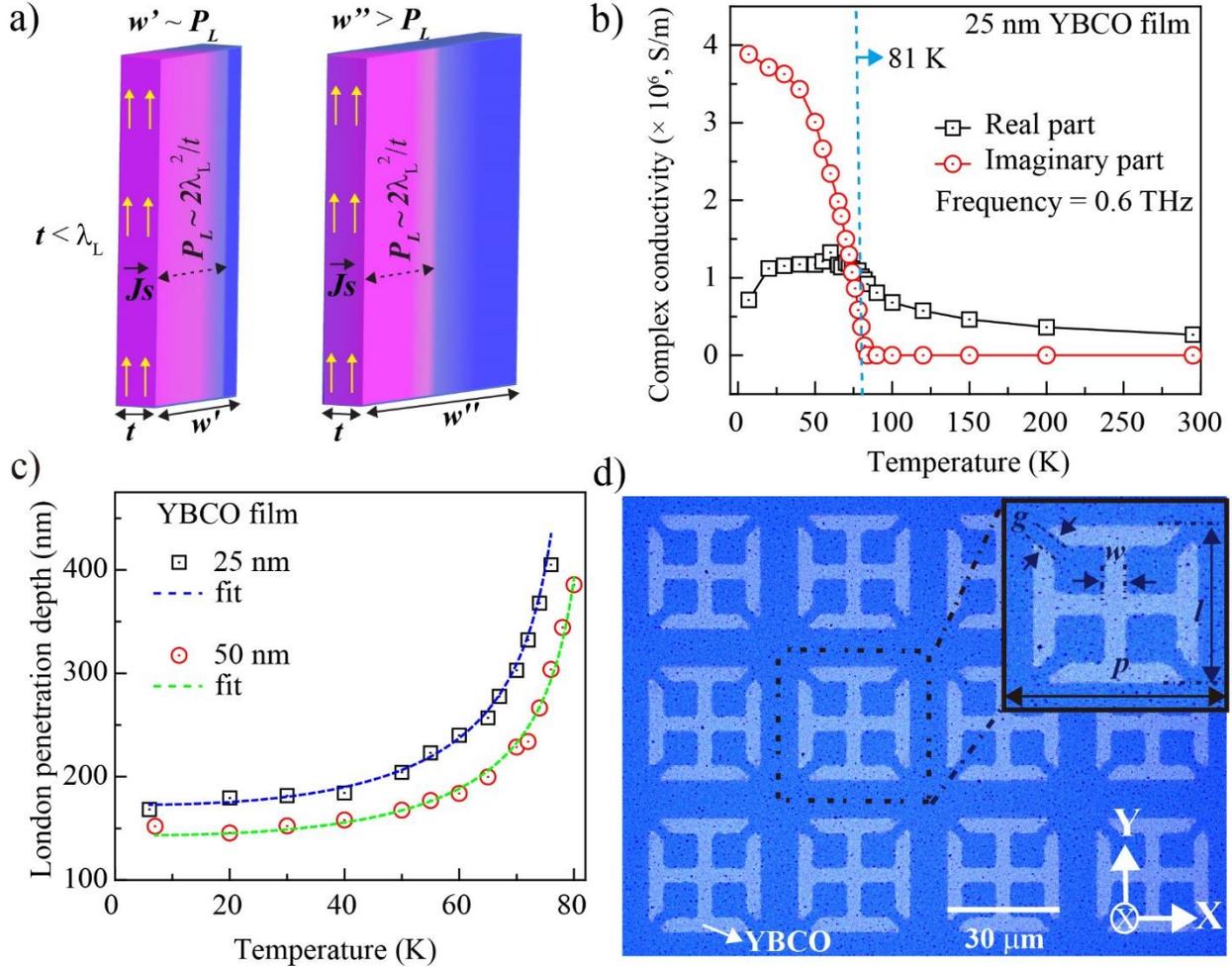

**Fig. 1: Pearl length in a thin current carrying superconductor film and Pearl metadevice design.** a) Schematic of thin current-carrying superconductor strips of width *w'*, comparable to Pearl length ($P_L$) and w", larger than $P_L$ showing current screening up to a length scale of $P_L$ (shown by pink region). b) Complex conductivity of 25 nm thin YBCO film at 0.6 THz obtained using terahertz transmission measurements in the temperature range 6 K to 300 K. c) Measured $\lambda_L$ of 25 nm and 50 nm YBCO film. The dashed curves represent the parametric fitting of the measured $\lambda_L$ with equation 1. d) Optical microscopy (OM) image of the fabricated 25 nm thin YBCO metadevice sample with the inset depicting the geometrical dimensions of a unit cell with resonator length, *l*: 30 µm; gap *g*: 4 µm, resonator width, *w*: 4 µm, and the periodicity, *p*: 40 µm.

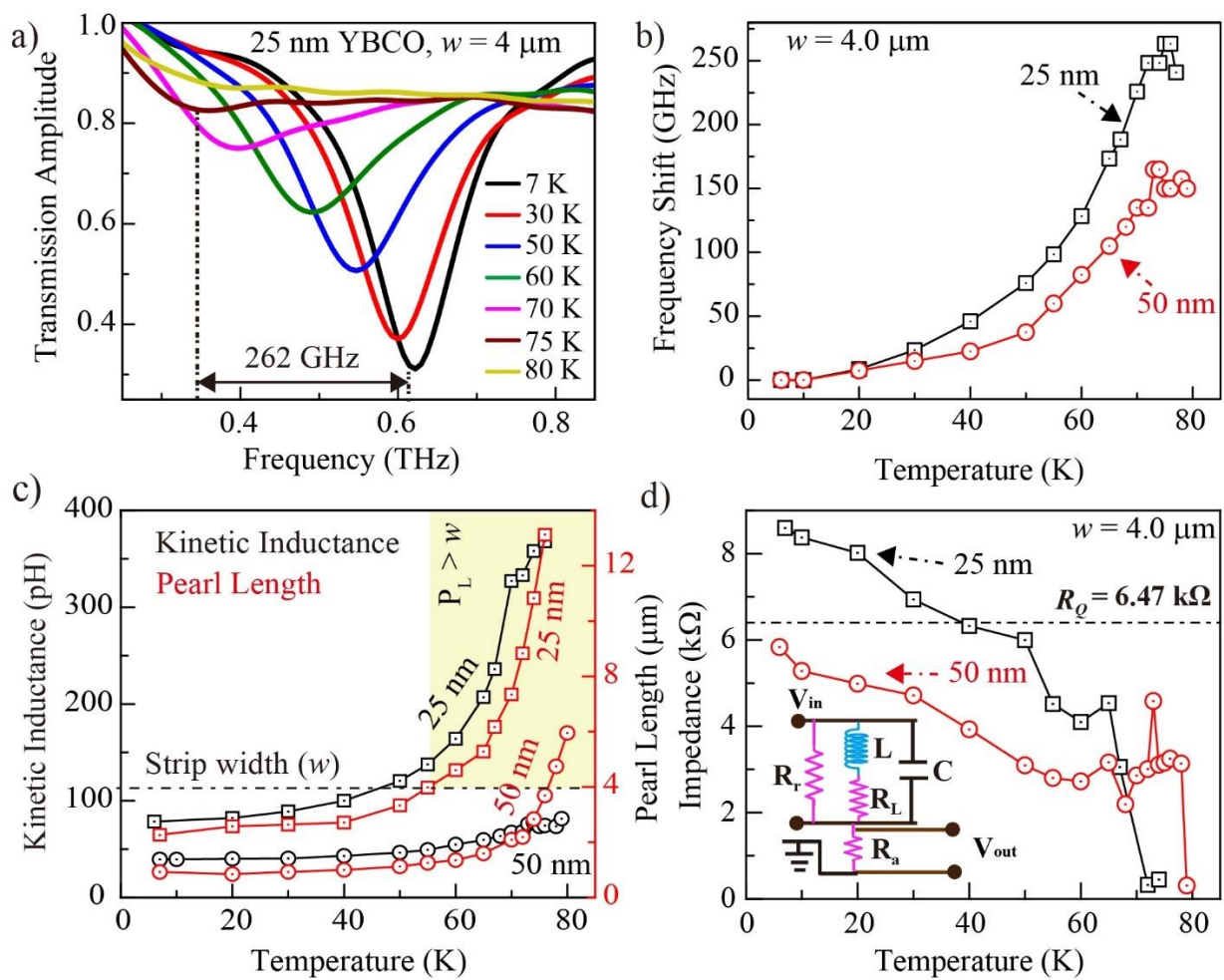

**Fig. 2: Giant Kinetic Inductance driven LC resonance shifts of the YBCO metadevice and equivalent circuit model**. a) Experimentally measured temperature-dependent transmission spectra of the 25 nm thin YBCO Pearl metadevice. b) Resonance frequency shift of 25 and 50 nm thin YBCO metadevice on varying temperature from 7 K to 80 K. c) Kinetic inductance obtained by circuit model fitting of measured transmission spectra of 25 and 50 nm thin YBCO metadevices and Pearl length obtained from the terahertz transmission measurement of 25 and 50 nm YBCO film, respectively. d) Extracted characteristic input impedance of 25 and 50 nm YBCO metadevice. Inset depicts designed equivalent transmission line-based circuit model.

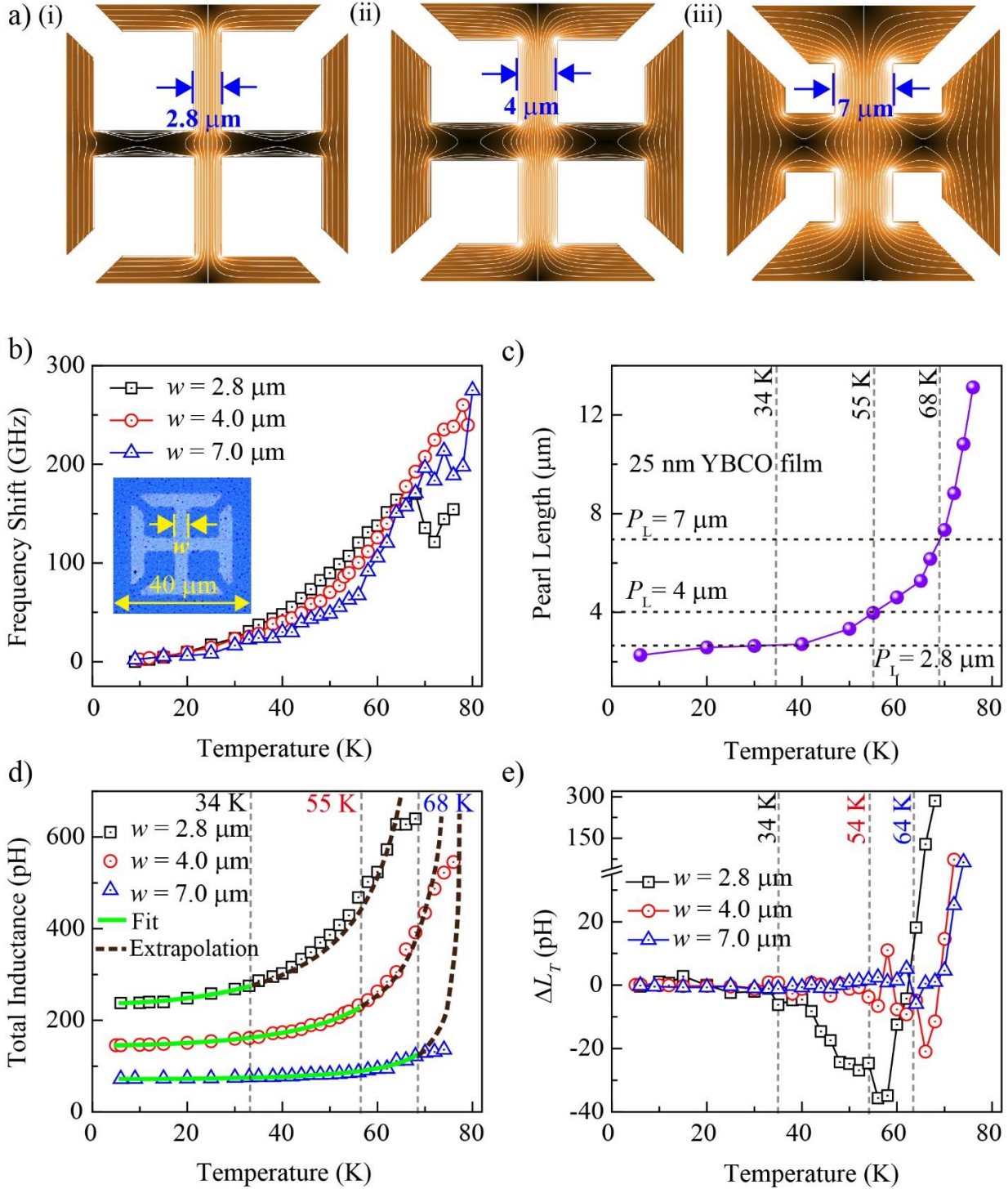

**Fig. 3: Role of Pearl length in establishing superinductance in thin superconductor metadevices.** a) Simulated current distribution in a unit cell of 25 nm YBCO Pearl metadevices with arm width $w$ of (i) 2.8 µm, (ii) 4 µm, and (iii) 7 µm. b) Resonance frequency shift obtained from the measured temperature-dependent transmission spectra of fabricated Pearl metadevices

with varying arm widths shown in (a). Inset shows the unit cell of the fabricated 25 nm YBCO Pearl metadevices having resonator width $w = 4$ μm. c) Pearl length of identical 25 nm YBCO thin unstructured film obtained through terahertz transmission measurements. Vertical black dotted lines represent the temperature at which the Pearl length exceeds the arm width w of corresponding Pearl metadevices. d) Temperature-dependent total inductance ($L_T$) of Pearl metadevices obtained by circuit model fitting of the measured transmission spectra at each temperature. Green solid and brown dotted curves represent the parametric fitting and extrapolation using parametric equation 2 (given in text) of $L_T$ derived for a superconductor strip of sub-London penetration depth thickness. e) $\Delta L_T$, difference between fitting values (solid and dotted curves in d) and calculated values (shown as scatter in d) of $L_T$.

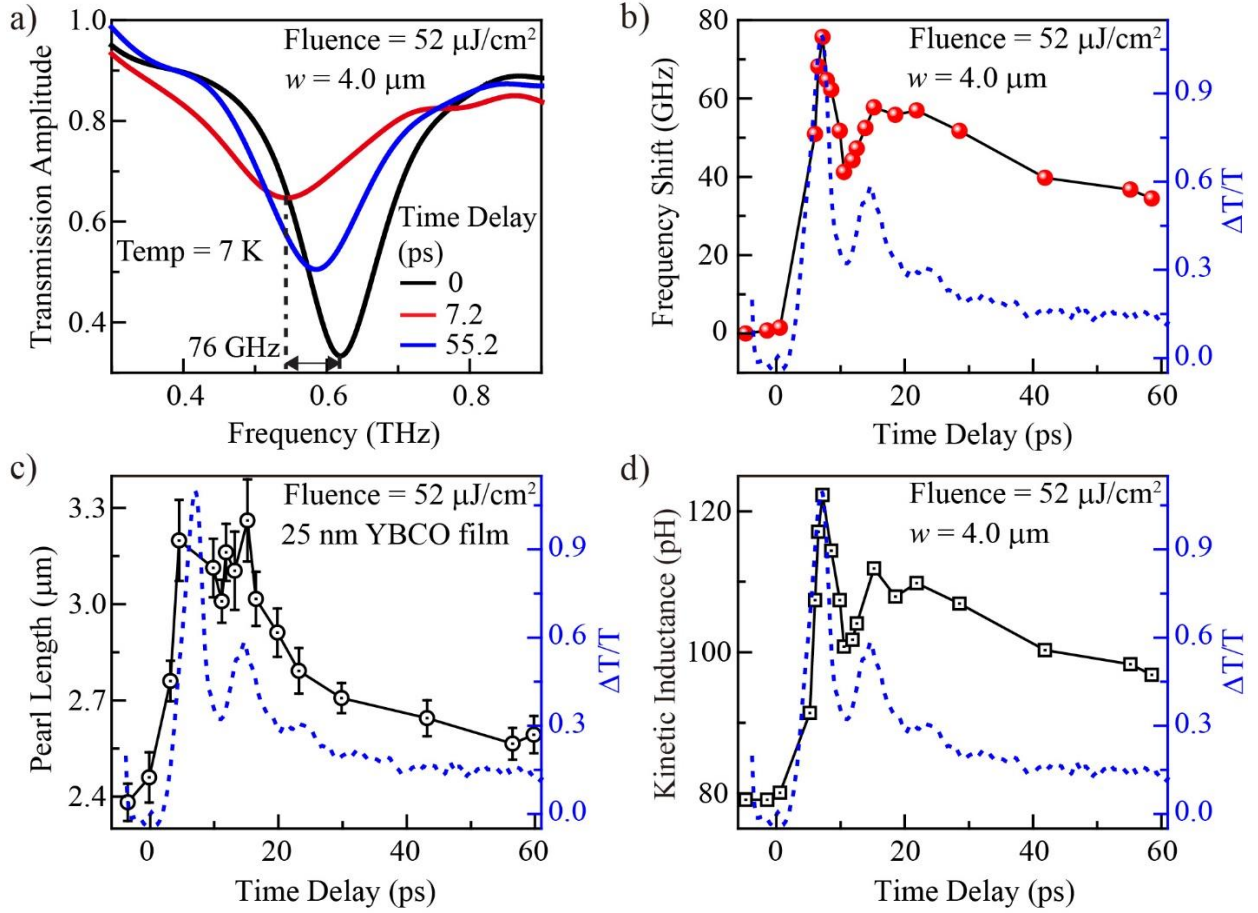

**Fig. 4: Ultrafast optical control of metadevice resonance frequency ($f$), Pearl length ($P_L$), and kinetic inductance ($L_K$).** a) The transmission spectra of 25 nm YBCO Pearl metadevice at 7 K on irradiation with an optical pump pulse of energy 1.55 eV and fluence of 52 µJ/cm², measured at varying time delay between optical pump and terahertz probe ($\tau_p$). b) The transient resonance frequency shift measured at varying $\tau_p$ corresponding to 0 ps time delay. c) Dynamic Pearl length of the 25 nm YBCO film and d) $L_K$ of the 25 nm YBCO Pearl metadevice measured at 7 K and varying $\tau_p$. The blue dashed curve in b), c), and d) depicts the measured Cooper pair dynamics at 7 K of the 25 nm YBCO thin film on irradiating it with an optical pump beam of fluence 52 µJ/cm².

**Acknowledgments**

The authors acknowledge research funding support from the Singapore Ministry of Education Academic Research Fund Grant Nos. MOE-Tier 2_EP50121-0009 and MOE2016-T3-1-006(S).

**Competing Interests Statement**

The authors declare no competing interests.